\documentclass[conference]{IEEEtran}
\IEEEoverridecommandlockouts
\usepackage{cite}
\usepackage{amsmath,amssymb,amsfonts}
\usepackage{algorithmic}
\usepackage{graphicx}
\usepackage{subfigure}
\usepackage{textcomp}
\usepackage{xcolor}
\usepackage{amsthm}
 
\theoremstyle{definition}
\newtheorem{definition}{Definition}[section]

\def\BibTeX{{\rm B\kern-.05em{\sc i\kern-.025em b}\kern-.08em
    T\kern-.1667em\lower.7ex\hbox{E}\kern-.125emX}}
\begin{document}

\title{A Novel Graph Analytic Approach to Monitor Terrorist Networks}

\author{\IEEEauthorblockN{Kaustav Basu, Chenyang Zhou, Arunabha Sen}
\IEEEauthorblockA{\textit{NetXT Lab, SCIDSE} \\
\textit{Arizona State University}\\
Tempe, Arizona \\
\{kaustav.basu, czhou24, asen\}@asu.edu}
\and
\IEEEauthorblockN{Victoria Horan Goliber}
\textit{Arizona State University}\\
Tempe, Arizona \\
victoria.goliber@gmail.com
}

\maketitle

\begin{abstract}
Terrorist attacks all across the world have become a major source of concern for almost all national governments. The United States Department of State's Bureau of Counter-Terrorism, maintains a list of 66 terrorist organizations spanning the entire world. Actively monitoring a large number of organizations and their members, require considerable amounts of resources on the part of law enforcement agencies. Oftentimes, the law enforcement agencies do not have adequate resources to monitor these organizations and their members effectively. On multiple incidences of terrorist attacks in recent times across Europe, it has been observed that the perpetrators of the attack were in the suspect databases of the law enforcement authorities, but weren't under active surveillance at the time of the attack, due to resource limitations on the part of the authorities. As the suspect databases in various countries are very large, and it takes significant amount of technical and human resources to monitor a suspect in the database, monitoring all the suspects in the database may be an impossible task. In this paper, we propose a novel terror network monitoring approach that will significantly reduce the resource requirement of law enforcement authorities, but still provide the capability of uniquely identifying a suspect in case the suspect becomes {\em active} in planning a terrorist attack. The approach relies on the assumption that, when an individual becomes active in planning a terrorist attack, his/her friends/associates will have some inkling of the individuals plan. Accordingly, even if the individual is not under active surveillance by the authorities, but the individual's friends/associates are, then {\em the individual planning the attack can be uniquely identified}. We apply our techniques on various real-world terror network datasets and show the effectiveness of our approach. 
\end{abstract}


\section{Introduction}
\label{Introduction}
Global terrorism is a major source of concern for national governments and law enforcement authorities all across the world, particularly in Europe and Asia. The Bureau of Counter-Terrorism, a division of the United States Department of State, maintains a list of 66 active terrorist organizations based all over the world. Actively monitoring such organizational networks is a considerable challenge on the part of the governments and law enforcement agencies in the sense that, significant amount of resources must be utilized to monitor these networks. Oftentimes, the law enforcement agencies do not have sufficient resources to monitor these organizations and their members effectively. It has been reported that the French Centre for the Analysis of Terrorism has determined that it takes as many as 20 agents per suspect to conduct 24-hour surveillance. On multiple incidences of terrorist attacks in recent times across Europe, it has been observed that the perpetrators of the attack were in the suspect databases of the law enforcement authorities, but weren't under active surveillance at the time of the attack due to resource limitations on the part of the authorities.



According to reports, the two suspects in the attack against French police on April 20, 2017, on the Champs-Elysees, were known to French anti-terrorism authorities. The November 2015 attacks in Paris that claimed a total of 130 lives, involved a small network of ISIS-linked terrorists in France and Belgium. Of the 10 individuals involved, several were known to authorities. When 12 people were killed at the Paris headquarters of Charlie Hebdo, a satirical magazine, all three of the terrorists had been under close watch. Cherif Kouachi, Said Kouachi and Amedy Coulibaly were under police surveillance for three years, but eventually dropped in the summer of 2014 only months before the deadly January 2015 attack \cite{wiki}.

As the suspect databases in various countries are very large, and it takes significant amount of technical and human resources to monitor a suspect in the database, monitoring all the suspects in the database may be an impossible task. 
The news organization Politico, \cite{politico} reported in October 2016 that the French authorities were monitoring around 15,000 individuals who were suspected of being radical Islamists. The Politico report was based on an earlier publication in the French journal, La Journal du Dimanche. The ABC news affiliated TV station WJLA in Washington D.C., reported in 2017 that, the list has tripled over the last two years \cite{wjla}. The database is managed by France’s Counter-Terrorism Coordination Unit. Obviously, the resources and manpower needed to keep all the terror suspects under surveillance is enormous and often are way beyond the available resources of any local law enforcement authority.

In this paper, we propose a novel Terrorist Network Monitoring (TNM) approach that will significantly reduce the resource requirement of law enforcement authorities, but still provide the capability of uniquely identifying a suspect in case the suspect becomes {\em active} in planning a terrorist attack. The approach relies on the assumption that, when an individual becomes active in planning a terrorist attack, his/her friends/associates will have some inkling of the individuals plan. Accordingly, even if the individual is not under active surveillance by the authorities, but the individual's friends/associates are, then {\em the individual planning the attack can be uniquely identified}. The mathematical foundation of our approach relies on {\em Identifying Code} and its variation known as {\em Discriminating Code} and is described in detail in Section \ref{Codes}. Our Identifying Code based approach ensures that the resource requirement of the law enforcement authorities will be significantly reduced, without compromising the ability of unique identification of a suspect in case he/she becomes {\em active} in planning a terrorist attack. 
 
We study terrorist networks from two perspectives. In the first perspective, the nodes represent individuals and the edges represent their relationships. In the second, we have two different types of nodes, one representing individuals and the other representing the organizations. The edges in this version do not represent individual to individual relationships, instead they represent the relationships between individuals and organizations. Our approach utilizes Identifying Codes for the study of the first type of networks, and Discriminating Codes for the second type. We apply our techniques on six real-world terror network datasets and show the effectiveness of our approach. The networks for our analysis, were obtained from the UCINET repository online \cite{ucinet}. In the following paragraphs, we briefly discuss our datasets. 

The first network on which we applied our technique is the network of the individuals involved in the terror attack in Paris in November 2015. This network, as shown in Fig. \ref{fig:Par}, had 10 nodes and 14 edges. Our analysis showed that, if 5 of the 10 individuals were monitored, the attackers most likely would have been exposed. 

We next analyze the Rizal Day bombings in 2000, where a series of bombs exploded in Metro Manila, Philippines. Our analysis technique requires that neighborhood of each node of the network be \textit{unique}. The reason for such a requirement is explained in Section \ref{Codes}. In this network, 16 individuals were involved, but two sets of two individuals had identical neighborhoods. In order to satisfy our analysis technique requirement, that neighborhood of each node of the network be \textit{unique}, we combined nodes with identical neighborhoods into a single \textit{super-node}. After the combination, the network has 14 nodes and 52 edges, and is shown in Fig. \ref{fig:Phi}. Our analysis showed that the attackers could have been exposed if 6 of the 14 individuals (nodes) were monitored. 

The IS-Europe network (or the Zerkani Network), shown in Fig. \ref{fig:IS}, forms the third network in our study. For the same reason as in the Philippines network, a pair of nodes were combined to form a \textit{super-node}. After the combination, the network had 39 nodes. We show that if 17 of the 39 nodes were monitored, then the attackers would have been exposed.

The Madrid train bombings network of 2004, shown in Fig. \ref{fig:Mad}, is the fourth network in our analysis. Out of the 54 individuals involved the bombings, if 17 were monitored, then the attackers could have been identified. 

The fifth network is the Noordin Mohammed Top network \cite{liebig}. The authors in \cite{liebig} identified four individuals, Noordin Top, Azhari Husin, Purnama Putra and Ahmad Ridho, as the key individuals of the network. In our analysis, we view these four individuals as layer 1 individuals, and the remaining as layer 2 individuals. As monitoring the key players (layer 1 individuals) is often significantly more difficult than monitoring layer 2 individuals, we focus on monitoring the layer 1 individuals in an indirect manner, by directly monitoring layer 2 individuals who are interacting with the layer 1 individuals. To capture this mechanism, we construct a bipartite graph $G = (V_1 \cup V_2, E)$, where the nodes in $V_1$ represent the individuals in layer 1 and the nodes in $V_2$ represent the individuals in layer 2. An edge $e \in E$ connects a node $u \in V_1$ with a node $v \in V_2$, if $u$ interacts with $v$. In this network $|V_1| = 4$ and $|V_2| = 15$, and is shown in Fig. \ref{fig:NT}. Our analysis shows that, if 3 of the 15 layer 2 individuals were monitored, then all four layer 1 individuals could have been exposed. 

The final network in our analysis is a network of terrorist organizations and their members, based in East Turkestan. This is a  bipartite graph $G = (V_1 \cup V_2, E)$, where $V_1$ represents the organizations and $V_2$ represents the individuals. An edge $e \in E$ connects a node $u \in V_1$ with a node $v \in V_2$, if $v$ is a member of $u$. This network is shown in Fig. \ref{fig:BP}. We show that if 15 out of 64 terrorists were monitored then the 20 terrorist organizations would have been uniquely monitored.    

The remainder of the paper is organized as follows. In Section \ref{Related Work}, we review the literature related to terrorist network analysis as well as Identifying Codes. Section \ref{Codes} is an overview of the mathematical concepts of Identifying Codes and Discriminating Codes. The problem of monitoring TNM is formally presented in Section \ref{Problem Formulation}. Section \ref{Problem Solution} provides the solution technique to the terrorist monitoring problem. In Section \ref{Experimental Results} we present the results of our experiments on the six real world terror networks described earlier. Section \ref{Conclusion} concludes the paper.

\section{Related Work}
\label{Related Work}

In the past few years, significant research on counter-terrorism has been conducted through social network analysis. In this section, we highlight a few key contributions that forms the basis for this effort. The authors in \cite{ressler}, investigate the suitability of social network analysis for studying terrorist networks. Krebs in \cite{krebs}, mapped the 9/11 terror network from articles in leading newspapers. Carley in \cite{carley}, explored the potential of using social network analysis and multi-agent modeling for the purpose of destabilizing terrorist networks. Julei in \cite{julei}, studied the bipartite networks of terrorist organizations in East Turkestan. 

In an earlier effort, we introduced the concept of utilizing Identifying Codes for monitoring terrorist networks in \cite{sen}. Our current effort expands on the results in \cite{sen} in multiple directions. In our earlier effort, we analyzed individual to individual networks to obtain the unique signature for each individual (node) in the network. In \cite{sen}, we did not consider the scenario of multiple individuals becoming active simultaneously. If multiple individuals (say, 2) become active simultaneously, not only do we need unique signature for each individual node, but also need unique signature for every pair of nodes. We consider this scenario in our current effort. Moreover, we consider individual to organization type of networks, which was not considered in \cite{sen}. Formalism of this scenario gives rise to computation of Discriminating Code for a bipartite graph formed by organizations and individuals. Discriminating Codes were not studied in \cite{sen}. Finally, in this paper we present the results of our experimentation on a much larger dataset than what was presented in \cite{sen}.
 
In addition to counter-terrorism research through social network analysis, the last few years have seen a significant amount of research on Identifying Codes and its applications in networks. Karpovsky {\em et. al.} \cite{karpov98} introduced the concept of Identifying Codes in  \cite{karpov98} and  provided results for Identifying Codes for graphs with specific topologies, such as binary cubes and trees. 
Using Identifying Codes, Laifenfeld {\em et. al.} studied joint monitoring and routing in wireless sensor networks in \cite{laifen09}.  Charon {\em et. al.} in \cite{charon2003}, studied complexity issues related to computation of minimum Identifying Codes for graphs and showed that in several types of graphs, the problem is NP-hard. Ray {\em et. al.} in \cite{ray03} generalized the concept of Identifying Codes, to incorporate robustness properties to deal with faults in sensor networks.

A special case, where only a subset of nodes needs a unique code, can be modeled with a bipartite graph, and this version of Identifying Codes is called ``Discriminating Codes'' and was studied in \cite{Discriminating}. This special case is relevant for our study as, our problem formulation of individual to organization network requires us to find the unique signatures of all nodes of one side of bi-partition of a bipartite graph, by selecting only a subset of the nodes in the other side of bi-partition. This formulation corresponds directly to ``Discriminating Codes''.

\section{A Brief Summary of Identifying and Discriminating Codes}
\label{Codes}

In this section, we formally define the mathematical notion of \textit{Identifying and Discriminating Codes}. Subsequently, we present our novel approach which utilizes these notions to uniquely monitor a given terror network.

\begin{definition}
Given a graph $G = (V, E)$, the subset $V' \subseteq V$, is defined as an Identifying Code Set (ICS) for the vertex set $V$, if $\forall v \in V, N[v] \cap V'$ is unique, where, $N[v] = v \cup N(v)$ and $N(v)$ represents the set of nodes adjacent to $v$ in $G = (V, E)$. The Minimum Identifying Code Set (MICS) problem is to find the ICS of smallest cardinality.  

\end{definition}

\begin{figure}[t]
\centering
	\includegraphics[scale=.40]{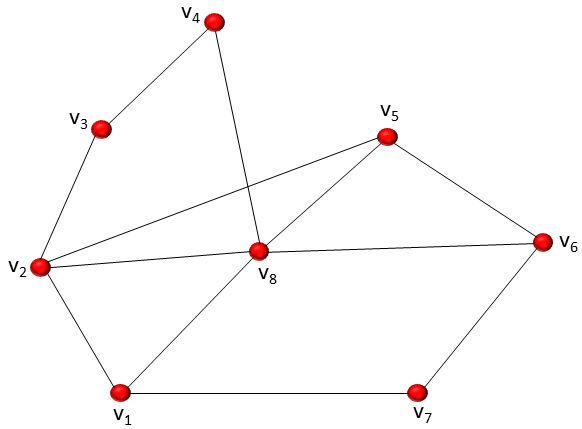}
	%
	%
	\caption{Graph with Identifying Code Set $\{v_4, v_6, v_7, v_8\}$}
	\label{fig:1}       
\end{figure}

The vertices of the set $V'$ can be thought of as \textit{alphabets} of the code, and the string made up with the alphabets of $N[v]$ can be viewed as the unique ``code'' for the node $v$. This is better explained  with the help of the following example. Consider a graph $G = (V, E)$, as illustrated in Fig. \ref{fig:1}. The Identifying Code Set for $G$ is $V' = \{v_4, v_6, v_7, v_8\}$. As shown in Table \ref{firsttable}, in $G = (V, E)$, $\forall v \in V, N[v] \cap V'$ is unique. Hence, the node set $V' = \{v_4, v_6, v_7, v_8\}$ is an ICS of $G$. 

\begin{definition}
Two nodes $u, v \in V$ are said to be ``twins'' if $N[v] = N[u]$.
\end{definition}

\noindent\textbf{Observation:} Identifying Code Set (ICS) of a graph $G = (V, E)$ does not exist, if any two nodes $u, v \in V$ are ``twins''.


\begin{table}[ht]
	\centering
	\caption{$N[v] \cap V'$ results for all $v \in V$ for the graph in Fig. \ref{fig:1}}
	\vspace{-8.00pt}
	\begin{tabular} {| c  | c |}  \hline 
		$N[v_1] \cap V' = \{v_7, v8\}$ & $N[v_2] \cap V' = \{v_8\}$ \\ \hline
		$N[v_3] \cap V' = \{v_4\}$ & $N[v_4] \cap V' = \{v_4, v_8\}$ \\ \hline
		$N[v_5] \cap V' = \{v_6, v_8\}$ & $N[v_6] \cap V' = \{v_6, v_7, v_8\}$ \\ \hline
		$N[v_7] \cap V' = \{v_6, v_7\}$ & $N[v_8] \cap V' = \{v_4, v_6, v_7, v_8\}$ \\ \hline
	\end{tabular}

\label{firsttable}
\end{table}

\begin{definition}
Given a bipartite graph $G = (V_1 \cup V_2, E)$, the subset $V'_2 \subseteq V_2$, is defined as a Discriminating Code Set (DCS) for the vertex set $V_1$, if $\forall v \in V_1, N(v) \cap V'_2$ is unique where, $N(v) \subseteq V_2$ represents the set of nodes adjacent to $v \in V_1$. The Minimum Discriminating Code Set (MDCS) problem is to find the DCS of smallest cardinality.  
\end{definition}

\begin{figure}[t]
\centering
	\includegraphics[scale=.40]{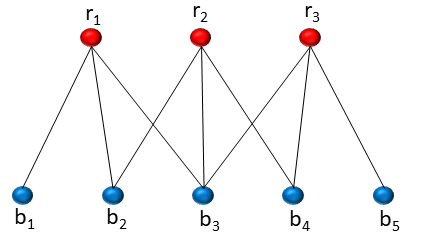}
	%
	%
	\caption{Bipartite Graph with Discriminating Code Set $\{b_2, b_4\}$}
	\label{fig:2}       
\end{figure}

Thus, the Discriminating Code problem is a restricted version of the Identifying Code problem, where the graph is bipartite. Simply stated, the monitoring problem in such a scenario, is to obtain unique signatures for nodes in $V_1$ using a subset of the nodes in $V_2$. We illustrate this with the help of the an example, shown in Fig. \ref{fig:2}. Here, $V_1 = \{r_1, r_2, r_3\}$, $V_2 = \{b_1, b_2, b_3, b_4, b_5\}$ and the DCS $V'_2 = \{b_2, b_4\}$, as $\forall v \in V_1, N(v) \cap V'_2$ is unique, as shown in Table \ref{secondtable}. 

\noindent\textbf{Observation:} Discriminating Code Set (DCS) of a bipartite graph $G = (V_1 \cup V_2, E)$ does not exist if any two nodes $u, v \in V_1$ are ``twins''. 

\begin{table}[h!]
	\centering
	\caption{$N(v) \cap V'_2$ results for all $v \in V_1$ for the graph in Fig. \ref{fig:2}}
	\vspace{-8.00pt}
	\begin{tabular} {| c |}  \hline 
		$N(r_1) \cap V'_2 = \{b_2\}$ \\ \hline
		$N(r_2) \cap V'_2 = \{b_2, b_4\}$ \\ \hline
		$N(r_3) \cap V'_2 = \{b_4\}$  \\ \hline
	\end{tabular}

\label{secondtable}
\end{table}
  
In Section \ref{Introduction}, we indicated that in some networks, such as the Philippines and IS-Europe, we combined multiple nodes into a single \textit{super-node}. The reason for this combination is the following. We noted that neither Identifying Code nor Discriminating Code can be computed if the corresponding graphs have ``twins''. In the Philippine network, the nodes 6 and 7 as well as 14 and 15 are ``twins''. In the IS-Europe network, the nodes 33 and 35 are ``twins''. By combining the ``twin'' nodes into a \textit{super-node}, we can ensure computation of Identifying and/or Discriminating Code in the modified network. However, as the modified network does not have node 6 or 7 (it has a super-node (6, 7)), if the individual corresponding to node 6 or 7 becomes active, Identifying Code will not be able to distinguish between these two individuals. If there is any indication that the super-node (6, 7) is in the process of being active, then further (lower level) analysis will be needed to find out whether node 6 or node 7 is in the process of being active. \\

\noindent
{\tt MICS and MDCS computation as a Graph Coloring with Seepage (GCS) Problem}:\\ The MICS and MDCS computation problem can be viewed as a novel variation of the standard Graph Coloring problem. We will refer to this version as the {\em Graph Coloring with Seepage (GCS)} problem. In the standard graph coloring problem, when a color is {\em assigned} (or injected) to a node, only that node is colored. The goal of the standard graph coloring problem to use as few distinct colors as possible such that (i) every node receives a color, and (ii) no two adjacent nodes of the graph have the same color. In the GCS problem, when a color is assigned (or injected) to a node, not only that node receives the color, the color also {\em seeps} into all the adjoining nodes. As a node $v_i$ may be adjacent to two other nodes  $v_j$ and  $v_k$ in the graph, if the color red is injected to $v_j$, not only will $v_j$ become red, but also $v_i$ will become red as it is adjacent to  $v_j$. Now if the color blue is injected to $v_k$, not only will $v_k$ become red, but also, the color blue will seep in to $v_i$ as it is adjacent  $v_k$. Since $v_i$ was ready colored red (due to seepage from $v_j$), after color seepage from $v_k$, its color will be a {\em combination of red and blue, i.e., purple}. At this point all three nodes  $v_i$, $v_j$ and $v_k$ have a color and all of them have distinct colors (red, blue and purple). The goal of the GCS problem is to inject colors to as few nodes as possible, such that (i) every node receives a color, and (ii) no two nodes of the graph have the same color.\\
Suppose that the node set $V'$ is an ICS of of a graph $G = (V, E)$ and $|V'| = p$.  In this case if $p$ distinct colors are injected to the nodes of  $V'$ (one distinct color to one node of $V'$ ), then as by the definition of ICS for all $v \in V$ if $N[v] \cap V'$ is unique, all nodes of  $G = (V, E)$ will be colored and no two nodes will have the same color. Accordingly, computation of the MICS problem is equivalent to computation of the GCS problem.  

Similar argument also holds for computation of the MDCS problem. Assume that the node set $V_2'$ is a DCS of of a graph $G = (V_1 \cup V_2, E)$ and $|V'_2| = p$.  In this case if $p$ distinct colors are injected to the nodes of  $V'_2$ (one distinct color to one node of $V'_2$ ), then as by the definition of DCS for all $v \in V_1$ if $N(v) \cap V'_2$ is unique, all nodes $v \in V_1$ will be colored and no two nodes will have the same color.

\section{Problem Formulation}
\label{Problem Formulation}

In this section, we formalize the Terrorist Network Monitoring problem using two different types of networks. In the first type of network, the nodes represents individuals (terrorists) and the edges represent the relationships between the individuals (may be friends/associates). In the second type, we have two different types of nodes - organizations and individuals. The edges of this type represent the relationship between the individuals and the organizations. In the following two subsections, we elaborate on our study of these two different types of networks.

\subsection{Individual to Individual (I-to-I) Network}

In this section, we formalize three different versions of the monitoring problem on individual to individual networks.

\noindent\textbf{Version 1:} If the network is represented as a graph $G = (V, E)$, the goal of the first version is to monitor\footnote{A monitor in this context means surveillance of the individual} a subset of nodes $V' \subseteq V$, so that every node $v \in V$ can be \textit{uniquely} identified, in case the individual representing the node $v$ becomes active in planning an attack. This version assumes that only one node becomes active at a time (i.e., simultaneous activation of two or more nodes is not considered in this version). Moreover, in this version, any node $v \in V$ can be monitored. 

The goal of this version of the TMN problem is to monitor the fewest number of terrorists (through surveillance) so that, if any one terrorist becomes active (planning an attack) in the network, then this individual can be uniquely identified. As discussed in Section \ref{Codes}, the goal of the GCS problem is to inject colors to as few nodes as possible, so that (i) every node receives a color, and (ii) no two nodes of the graph have the same color. The GCS problem has direct correspondence with the TNM problem, in the sense that, the terrorists to be monitored in the TNM problem, may be viewed as the nodes where colors have to be injected in the GCS problem. As injecting colors into this set of nodes will ensure that each node of the network will have a unique color, monitoring the corresponding terrorists will provide us with a unique signature of each node (terrorist) in the network. In other words, unique signature of a terrorist will correspond to the unique color associated with  the corresponding node. As further discussed in Section \ref{Codes}, the GCS problem is equivalent to the computation of the MICS problem. By transitivity, it is clear that the TNM problem is equivalent to solving the MICS problem of the corresponding terrorist network. 

\noindent\textbf{Version 2:} This version is almost identical to Version 1, except that in this version, simultaneous activation of at most two nodes is allowed. In version 1, we required that each node has its unique signature. The version 1 excludes the possibility of more than one terrorist becoming active at the same time. In this version, we relax this constraint and consider the possibility of at most two terrorists becoming active simultaneously. In this scenario, not only we require that every terrorists has a unique signature (i.e., every node has a unique color), but also every pair of terrorists (nodes) has a unique signature (color). Thus, this version of the TNM problem is equivalent to solving the MICS problem with this additional constraint.     

\noindent\textbf{Version 3:} The versions 1 and 2 assume that any node $v \in V$, may become active and can be monitored. In this version, only a subset $V_1 \subset V$ needs to be monitored and monitors can only be placed in the subset $V_2 = V - V_1$. In this case, we can construct a bipartite graph $G = (V_1 \cup V_2, E)$, where each node in $V_1$ is required to have a unique color. The goal of this version of the problem is to inject colors at the fewest number of nodes in $V_2$, such that every node in $V_1$ receives a unique color. Thus, this version of the TNM problem is equivalent to solving the MDCS problem, discussed in Section \ref{Codes}.

\subsection{Individual to Organization (I-to-O) Network}

In this subsection, we study individual to organizational networks. If the network is represented as a graph $G = (V_1 \cup V_2, E)$, where $V_1$ represents the organizations and $V_2$ represents the individuals. There is an edge between a node $u \in V_1$ to a node $v \in V_2$ if the individual represented by $v$ is a member of the organization represented by $u$. The goal of this version is to \textit{uniquely} identify an organization (if the organization becomes active), by monitoring a subset of individuals belonging to these organizations. Accordingly, this version of the TNM problem is  equivalent to solving the MDCS problem.

\section{Problem Solution}
\label{Problem Solution}
In this section, we provide solution techniques for three versions of I-to-I networks and one version of I-to-O network, using Integer Linear Programs (ILP). In the following, we provide the ILP formulations for each of the versions. 
\subsection{I-to-I Network: Version 1}
\noindent{\textit{Instance:}} $G = (V, E)$, an undirected graph.\\
\textit{Problem}: Find the smallest subset $V' \subseteq V$, such that injection of colors at these nodes, ensures that each node $v \in V$, receives a unique color (either atomic or composite) through seepage.

\noindent {We use the notation $N[v_i]$ to denote the closed neighborhood of $v_i$, for any $v_i \in V$.} Corresponding to each $v_i \in V$, we use an indicator variable $x_i$,
\[x_i = \left\{ \begin{array}{ll}
                    1, & \mbox{if a color is injected at node $v_i$, } \\
                    0, & \mbox{otherwise}
                    \end{array}
            \right. \]

\noindent\textit{Objective Function:}  \textit{Minimize} $\sum_{v_i \in V}x_i$   \\ \\ 
\textit{Coloring Constraint:}  $\sum_{v_i \in N[v_j]}x_i \geq 1,$  $\forall v_j \in V$ \\ \\
\textit{Unique Coloring Constraint:}

$\sum_{v_i \in \{N[v_j] \bigoplus N[v_k]\} }x_i \geq 1,$ $\forall v_j \neq v_k, \in V$ \\

\noindent{$N[v_j] \bigoplus N[v_k]$ denotes the Exclusive-OR of the node sets $N[v_j]$ and $N[v_k]$.}
It may be noted that the objective function ensures that the fewest number of nodes in $V$ are assigned a color. The Coloring Constraint ensures that every node in $V$ receives at least one color through seepage from the colors injected at nodes in its closed neighborhood. A consequence of the Coloring Constraint is that, a node in $V$ may receive more than one color through seepage from the colors injected at its neighborhood. The Unique Coloring Constraint ensures that, for every pair of nodes ($v_j, v_k$) in $V$, at least one node in the node set $N[v_j] \bigoplus N[v_k] \subseteq V$ is injected with a color. This guarantees that $v_j$ and $v_k$ will not receive identical colors.

\begin{figure*}[t]
\centering     
\subfigure[Paris Network]{\label{fig:Par}\includegraphics[width=65mm]{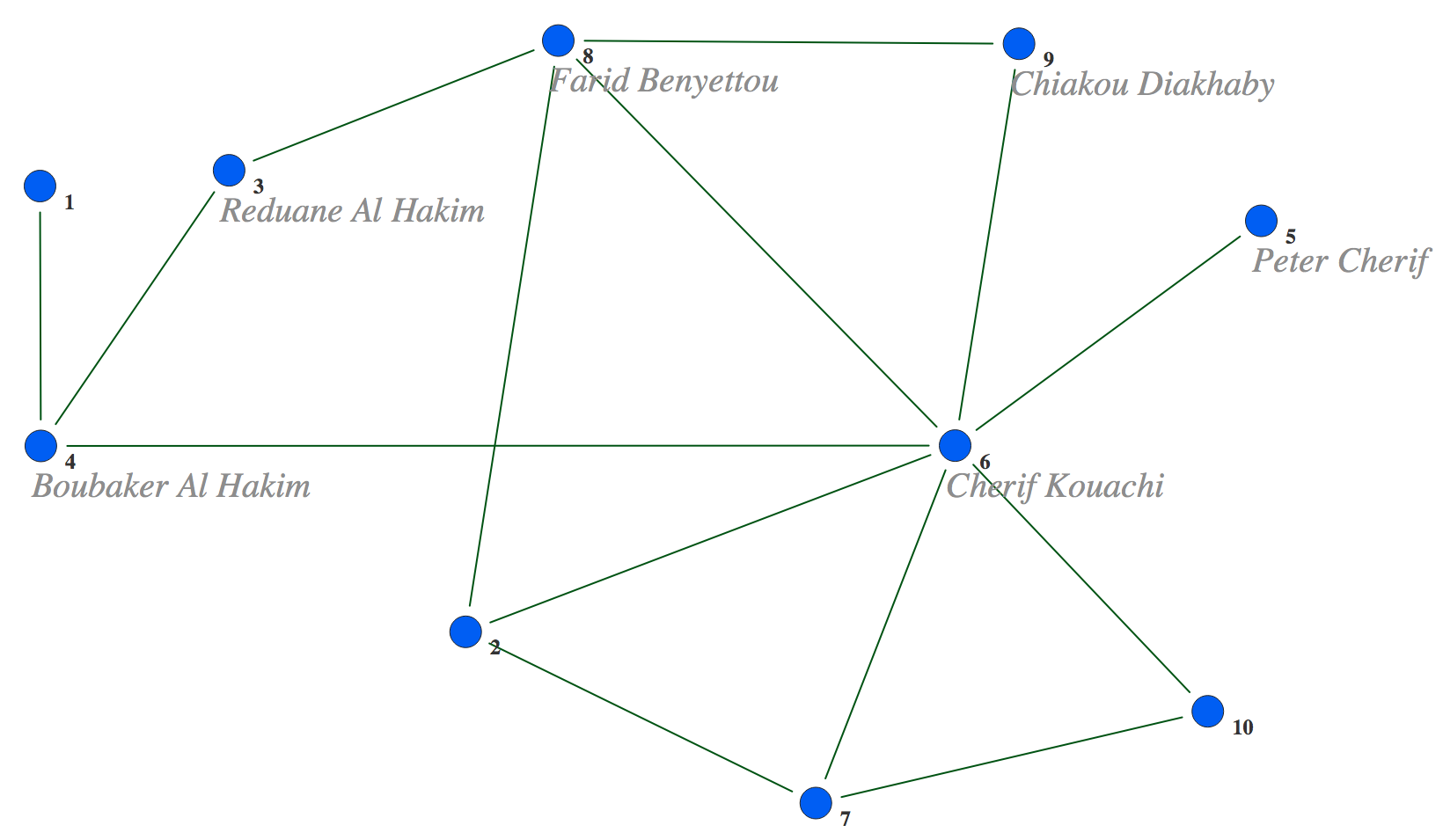}}
\subfigure[Philippines Network]{\label{fig:Phi}\includegraphics[width=65mm]{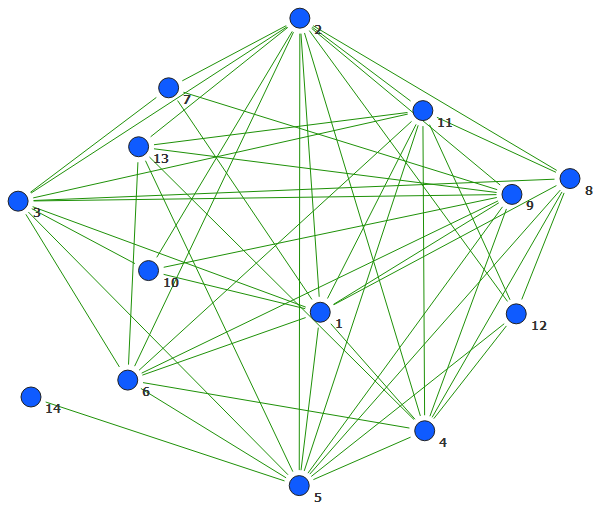}}
\caption{Terror networks involved in the 2015 Paris and 2000 Philippines attacks respectively}
\end{figure*}

\subsection{I-to-I Network: Version 2}
In this subsection we present an Integer Linear Program for determining the unique signatures for pairs of nodes. In this version, not only individual nodes are required to have unique signatures, but also every pair of nodes are also required to have unique signatures. This additional requirement (with respect to Version 1) necessitates that the ILP ensures that (i) a single node and a node pair do not end up having a identical signature, and (ii) two node pairs do not end up having a identical signature. Accordingly, the ILP is required to have four constraints: (i) coloring constraint (same as in version 1), (ii) unique coloring 1-1 constraint (same as unique coloring constraint in version 1), (iii) unique coloring 1-2 constraint (not present in version 1), and (iv) unique coloring 2-2 constraint (not present in version 1).

\noindent{\textit{Instance:}} $G = (V, E)$, an undirected graph.\\
\textit{Problem}: Find the smallest subset $V' \subseteq V$, such that injection of colors at these nodes, ensures that each node $v_i$ and each node pair $(v_j, v_k) \in V$, receive a unique color (either atomic or composite) through seepage.

\noindent {We use the notation $N[v_i]$ to denote the closed neighborhood of $v_i$, for any $v_i \in V$.} Corresponding to each $v_i \in V$, we use an indicator variable $x_i$,
\[x_i = \left\{ \begin{array}{ll}
                    1, & \mbox{if a color is injected at node $v_i$, } \\
                    0, & \mbox{otherwise}
                    \end{array}
            \right. \]

\noindent\textit{Objective Function:}  \textit{Minimize} $\sum_{v_i \in V}x_i$   \\ \\ 
\textit{Coloring Constraint:}  $\sum_{v_i \in N[v_j]}x_i \geq 1,$  $\forall v_j \in V$ \\ \\
\textit{Unique Coloring 1-1 Constraint:}

$\sum_{v_i \in \{N[v_j] \bigoplus N[v_k]\} }x_i \geq 1,$ $\forall v_j \neq v_k, \in V$ \\ 

\noindent\textit{Unique Coloring 1-2 Constraint:}

$\sum_{v_i \in \{N[v_j] \bigoplus (N[v_k] \cup N[v_l]\} )}x_i \geq 1,$ $\forall v_j \neq v_k \neq v_l, \in V$ \\ \\ \\

\noindent\textit{Unique Coloring 2-2 Constraint:}

\noindent $\sum_{v_i \in \{(N[v_j] \cup N[v_k]) \bigoplus (N[v_l] \cup N[v_m])\} }x_i \geq 1,$ $\forall v_j \neq v_k \neq v_l \neq v_m, \in V$ \\

\noindent The role of coloring constraint and unique coloring 1-1 constraint in this version is identical to the role of the coloring constraint and unique coloring constraint in version 1. These constraints ensure that no two nodes in the graph will have identical color. This version has two additional constraints. 

Color assigned to a pair of nodes $(v_k, v_l)$ is the union of the colors assigned to nodes $v_k$ and $v_l$. The Unique Coloring 1-2 Constraint ensures that, for every pair of ($v_j, (v_k, v_l)$) in $V$, at least one node in the node set $N[v_j] \bigoplus (N[v_k] \cup N[v_l]) \subseteq V$ is injected with a color. This guarantees that $v_j$ and $(v_k, v_l)$ will not receive identical colors. In other words, this constraint ensures that for all combinations of distinct nodes $u, v, w \in V$, the node $u$ will not have identical color as the node pair $(v, w)$. 

The Unique Coloring 2-2 Constraint ensures that, for every pair of ($(v_j, v_k), (v_l, v_m)$) in $V$, at least one node in the node set $(N[v_j] \cup N[v_k]) \bigoplus (N[v_l] \cup N[v_m]) \subseteq V$ is injected with a color. This guarantees that $(v_j, v_k)$ and $(v_l, v_m)$ will not receive identical colors. In other words, this constraint ensures that for all combinations of distinct nodes $u, v, w, x \in V$, the node pair $(u, v)$ will not have identical color as the node pair $(w, x)$. 

\begin{figure*}[t]
\centering     
\subfigure[IS-Europe Network]{\label{fig:IS}\includegraphics[width=90mm]{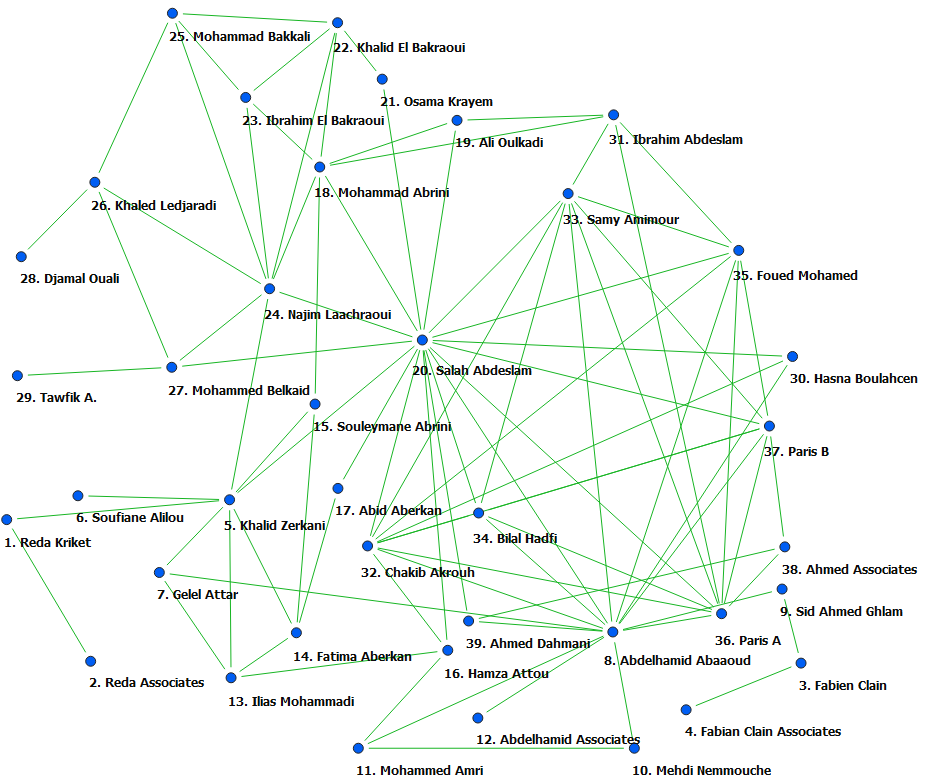}}
\subfigure[Madrid Network]{\label{fig:Mad}\includegraphics[width=80mm]{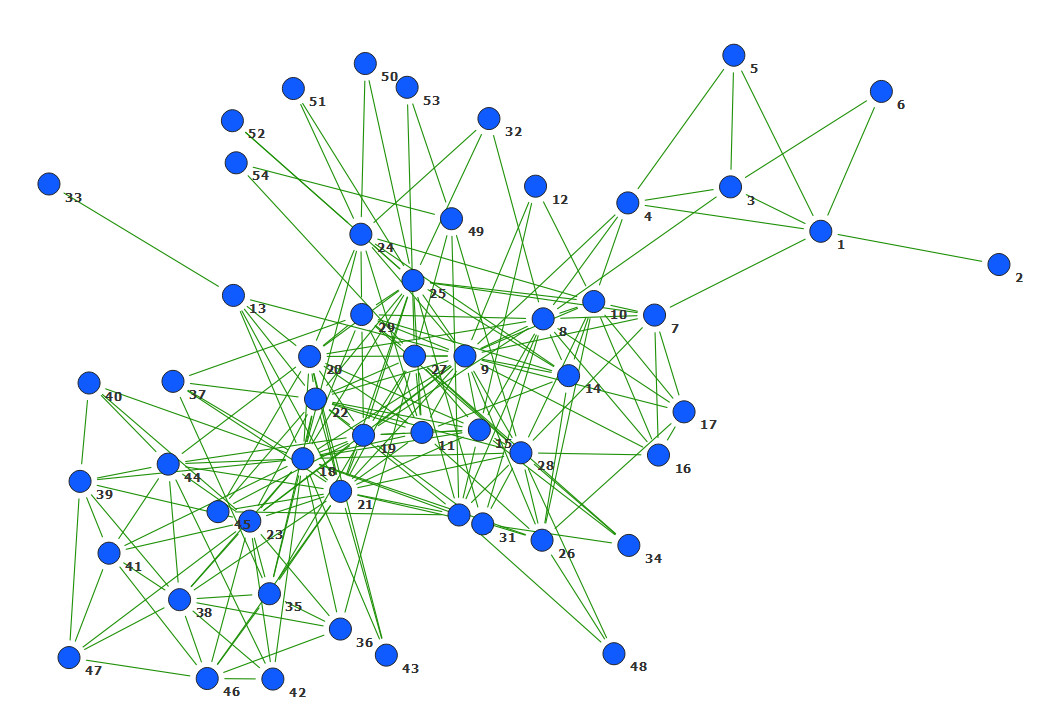}}
\caption{Terrorist Networks involved in the 2015 Paris/Brussels attacks and 2004 Madrid train bombings respectively}
\end{figure*}
\begin{figure*}[t]
\centering     
\subfigure[Noordin Top Key Player Network]{\label{fig:NT}\includegraphics[width=65mm]{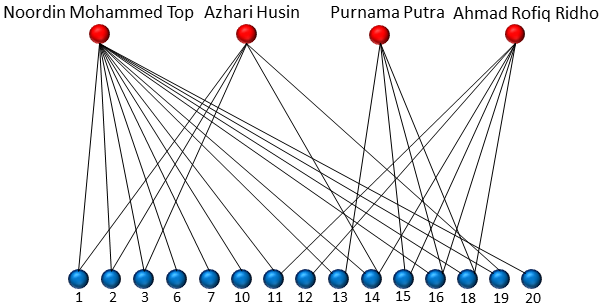}}
\subfigure[Terrorist-Terrorist Organization Network \cite{julei}]{\label{fig:BP}\includegraphics[width=65mm]{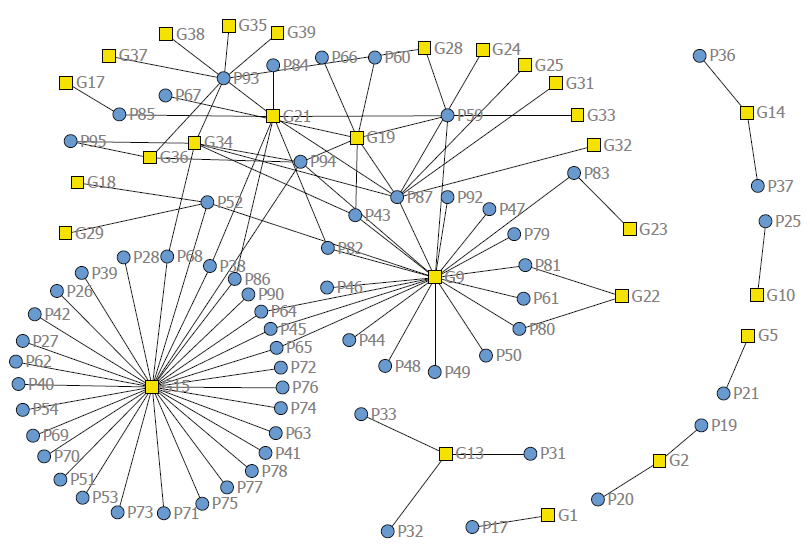}}
\caption{Terrorist networks involved in the 2004 Jakarta and East Turkestan attacks respectively}
\end{figure*}

\subsection{I-to-I Network: Version 3}
\label{DC}

It may be recalled that in this version, only a subset $V_1 \subset V$ needs to be monitored and monitors can only be placed in the subset $V_2 = V - V_1$. We construct a bipartite graph $G = (V_1 \cup V_2, E)$, where each node in $V_1$ is required to have a unique color. The goal of this version of the problem is to inject colors at the fewest number of nodes in $V_2$, such that every node in $V_1$ receives a unique color. 

\noindent{\textit{Instance:}} $G = (V_1 \cup V_2, E)$, an undirected bipartite graph.\\
\textit{Problem}: Find the smallest subset $V'_2 \subseteq V_2$, such that injection of colors at these nodes, ensures that $\forall v_i \in V_1$, receives a unique color (either atomic or composite) through seepage.

\noindent {We use the notation $N(v_i)$ to denote the neighborhood of $v_i$, for any $v_i \in V_1 \cup V_2$.} Corresponding to each $v_i \in V_2$, we use an indicator variable $x_i$,
\[x_i = \left\{ \begin{array}{ll}
                    1, & \mbox{if a color is injected at node $v_i$, } \\
                    0, & \mbox{otherwise}
                    \end{array}
            \right. \]

\noindent\textit{Objective Function:}  \textit{Minimize} $\sum_{v_i \in V_2}x_i$   \\ \\ 
\textit{Coloring Constraint:}  $\sum_{v_i \in N(v_j)}x_i \geq 1,$  $\forall v_j \in V_1$ \\ \\
\textit{Unique Coloring Constraint:}

$\sum_{v_i \in \{N(v_j) \bigoplus N(v_k)\} }x_i \geq 1,$ $\forall v_j \neq v_k, \in V_1$ \\

\noindent The notations and their explanations in this version is identical to the earlier two versions. It can be easily verified that the solution to the ILP finds the smallest subset $V'_2 \subseteq V_2$, such that injection of colors at these nodes, ensures that each node $v_i \in V_1$, receives a unique color. 

\subsection{I-to-O Network}
In both I-to-I network version 3 and I-to-O network, the input is a bipartite graph $G = (V_1 \cup V_2, E)$ and the goal is to find the smallest subset $V'_2 \subseteq V_2$ that assigns unique colors to all the nodes of $V_1$. Accordingly, the ILP formulation for the solution to these two problems is identical and hence not presented here. 

\section{Experimental Results}
\label{Experimental Results}

In this section, we present the results of our experimentations with six different datasets of real incidences. It may be noted that if the network under study $G = (V, E)$ is ``twin-free'', then the node set $V$ is an Identifying Code Set for the network, although it may not be the ICS of the smallest cardinality. The four networks shown in Table \ref{Version1}, have 10, 14, 39 and 54 nodes respectively. If monitors were placed on every node (i.e., colors were injected at every node) of the network, each node would have received a unique color (signature). However, our solution shows that unique signature for each node can be obtained by injecting colors at only 5, 6, 17 and 17 nodes respectively. As deploying a higher number of monitors (injecting colors) does not realize any additional benefit, there is no reason to deploy a larger number of monitors. The results presented in Table \ref{Version1} show that using our approach, resource requirement can be reduced by 50\%, 57.14\%, 56.41\% and 68.52\% respectively. As monitoring (surveillance) of a terrorist suspect involves significant cost on the part of law enforcement authorities, we expect that more than 50\% reduction of surveillance cost in every one of the four networks, will be of great interest to the authorities.     

In the following, we present the results in two subsections, for the I-to-I and for the I-to-O networks, respectively. 

\subsection{I-to-I Networks}

\noindent\textbf{Version 1:} In this version, we analyzed the Paris, Philippines, IS-E and the Madrid networks. These networks are illustrated in Figs. \ref{fig:Par}, \ref{fig:Phi}, \ref{fig:IS} and \ref{fig:Mad}. For ease of understanding, we have described the results of the Paris network in detail and tabulated the results of the other networks in Table \ref{Version1}.

\begin{table}
\centering
\caption{Version 1 Results}
\vspace{-8.00pt}
\begin{tabular}{| c | c | c | c |}
\hline 
 Network& Number of  &Colors Required & \% Reduction \\
&Nodes  & for Unique  & in Resources\\
&& Monitoring&\\ \hline
Paris Network & 10 & 5 & \textbf{50} \\ \hline
Philippines Network & 14 & 6 & \textbf{57.14} \\ \hline
IS-E Zerkani Network & 39 & 17 & \textbf{56.41}\\ \hline
Madrid Network & 54 & 17 & \textbf{68.52}\\ \hline
\end{tabular}
	\vspace{-12.00pt}
\label{Version1}
\end{table}

\begin{table}[h!]
	\centering
	\caption{Node color assignment in the Paris Network}
	\vspace{-8.00pt}
	\resizebox{\columnwidth}{!}{%
	\begin{tabular} {| c |c |c | c| | |l |c |l | c|}  \hline 
		Node & String & Node & String &String & Node & String & Node \\ \hline
		1 & B & 6 & ABCDE & ABCDE & 6 & BC & 4\\ \hline
		2 & ACDE & 7 & ACD & ACD & 7 & BE & 3\\ \hline
		3 & BE & 8 & ACE & ACDE & 2 & C & 5\\ \hline
		4 & BC & 9 &  CE & ACE & 8 & CD & 10\\ \hline
		5& C & 10 & CD & B & 1 & CE & 9\\ \hline
	\end{tabular}
	}
	\vspace{-12.00pt}
	\label{Paris}
\end{table}

\begin{table*}[h!]
\centering
\caption{Version 2 Results}
\vspace{-8.00pt}
\begin{tabular}{|c|c|c|c|c|}
\hline
Network & Number of  & Number of Colors & Number of Signatures Required for Unique & Number of Unique Signatures Produced\\
&Nodes & Injected  & Identification for Single Nodes and Node Pairs & After Injection of Colors at all Nodes \\ \hline
Paris & 10 & 10 & 55 & 42 \\ \hline
Philippines & 14 & 14 & 105 & 37 \\ \hline
Zerkani & 39 & 39 & 780 & 692 \\ \hline
Madrid & 54 & 54 & 1485 & 1155\\ \hline

\end{tabular}

\label{Version2}
\end{table*}

As indicated earlier, the Paris network consists of 10 individuals (nodes), who were involved in the attacks across multiple locations in Paris, in November 2015. The Identifying Code Set (ICS) for this network computed by the ILP given in Section \ref{Problem Solution}, is $V' = \{2, 4, 6, 7, 8\}$. This implies that, by injecting five colors $A, B, C, D, E$ to nodes in $V'$, all the nodes in $V$ receive a unique color (signature). This can be verified with the help of Table \ref{Paris}. This table illustrates the color string received by each node, and it is easy to verify that these color strings are unique. The results for the other three networks are presented in Table \ref{Version1}.

\noindent\textbf{Version 2:} In this version, we present the results of our approach when the simultaneous activation of two nodes is allowed. The ICS obtained for the Paris network, in this version, is $V' = \{1, 2, 3, 4, 5, 6, 7, 8, 9, 10\}$. In this network, there are 10 nodes (individuals) and $\binom{10}{2}$ = 45 possible ordering of node pairs. Thus to have unique signatures for single nodes and node pairs, there must be a total of 10 + 45 = 55 unique signatures. Our analysis for this network shows that, even if colors are injected in all 10 nodes of the network, it creates only 42 unique signatures. This happens because five node pairs ((2, 6), (5, 6), (6, 7), (6, 9) and (6, 10)) receive the same color. Moreover, there are four instances where two node pairs ((1, 8), (4, 8)), ((2, 7), (2, 10)), ((3, 6), (6, 8)) and ((7, 8), (8, 10)) produce the same signature. These 13 non-unique signatures are a result of the topology of the network. This implies that, in the Paris network, further analysis is needed to uniquely identify the node pairs that becomes active. The results for the other three networks are tabulated in Table \ref{Version2}.

\noindent\textbf{Version 3:} In this section, we analyzed the Noordin Mohammed Top network, shown in Fig. \ref{fig:NT}. The MDCS for this network is $V'_2 = \{7, 14, 18\}$, which implies that, four red nodes (corresponding to the four key individuals in this network) will receive a unique color if colors are injected in $V'_2$ only. Even if we assume that the cost of monitoring a key individual in a network is same as the cost of monitoring a low level individual (which most likely is false), monitoring three nodes (7, 14, 18) results in a \textbf{33.33\%} reduction in resource requirements, as compared to monitoring four red nodes.

\subsection{I-to-O Networks}

In this subsection, we analyzed the I-to-O network from East Turkestan, studied in \cite{julei}, and shown in Fig. \ref{fig:BP}. In this figure, the yellow squares (nodes) represent the terrorist organizations and the blue circles (nodes) represent the members of these organizations. It may be noted that this is a bipartite graph with a bi-partition formed by the blue and yellow nodes. This bipartite graph is not ``twin-free'', as several pairs of nodes such as G29 and G18 as well as G35 and G37 are ``twins''. In order to compute Discriminating Code Set of this bipartite graph, we first transformed it into a ``twin-free'' graph by combining several nodes into super-nodes. After the transformation, we have a bipartite graph with 20 yellow nodes (out of a total of 27, in the original graph). The MDCS for this transformed graph is $V'_2 = \{P94, P52, P85, P93, P60, P59, P87, P81, P83, P32, P17,$\\ $P19, P21, P25, P37 \}$. Using similar arguments as in I-to-I network version 3, it can be claimed that our Discriminating Code based technique results in a \textbf{25\%} reduction in resource requirements (15 monitors instead of 20 corresponding to each of the organizations).

\section{Conclusion}
\label{Conclusion}

In this paper, we presented a novel approach for monitoring terrorist networks that results in a significant reduction in resource requirements on the part of the law enforcement authorities. We considered four different scenarios and provided solution techniques for all of them. Moreover, we conducted extensive experimentation on six real world networks and demonstrated significant reduction in resource requirements.

\end{document}